Krzysztof Echaust, Krzysztof Piasecki[1]

# Black-Litterman model
# with intuitionistic fuzzy posterior return

**Abstract:** The main objective is to present a some variant of the Black - Litterman model. We consider the canonical case when priori return is determined by means such excess return from the CAPM market portfolio which is derived using reverse optimization method. Then the a priori return is at risk quantified uncertainty. On the side, intensive discussion shows that the experts' views are under knightian uncertainty. For this reason, we propose such variant of the Black - Litterman model in which the experts' views are described as intuitionistic fuzzy number. The existence of posterior return is proved for this case. We show that then posterior return is an intuitionistic fuzzy probabilistic set.

**Keywords:** Black-Litterman model, intuitionistic fuzzy number, quantitaive uncertainty, knightian uncertainty, imprecision

**JEL Classification:** C02 · G11

**Mathematics Subject Classification** 03E72 · 91G80

## 1. Research problem

The Black-Litterman model (BLM in the sequel) was introduced by Black and Litterman [6], expanded in [7,8] and discussed in detail in [3, 11, 16, 27]. BLM combines the CAPM [23], reverse optimization [24], mixed estimation [25, 26], the universal hedge ratio from Black's global CAPM [4, 5, 16], and mean-variance optimization [18]. The BML is applied for asset allocation in many financial institutions. This model provides the flexibility of combining the market equilibrium with additional market views of the investor.

In subject literature we have many versions of BLM. In each version investor's views are represented by vector of random variables. This representation requires the assumption that investor's views are under quantitative uncertainty. On the other side, this assumption is not empirically verifiable because of the investor's views are very intuitive. Thus, we can only assume that investor's views are under knightian uncertainty [14].

---

[1] Krzysztof Echaust
k.echaust@ue.poznan.pl

✉ Krzysztof Piasecki
k.piasecki@ue.poznan.pl

Deparment of Operations Research, Poznań University of Economics , al. Niepodległości 10, 61-875 Poznań, Poland

The intuitionistic fuzzy sets [1] may be applied as an image of knightian uncertainty. Therefore, the main aim of this article is to present possibility of using the intuitionistic fuzzy sets to describe investor's views.

## 2. Black-Litterman model - the basic case

BML uses the Bayesian approach to infer the assets' expected returns [6]. With the Bayesian approach, the expected returns are random variables themselves. They are not observable. One can only infer their probability distribution. The inference starts with a prior belief. Additional information is used along with the prior to infer the posterior distribution. In BLM, the prior distribution is the CAPM equilibrium distribution and the investor's views are the additional information.

The set $\Omega$ is a set of all elementary states $\omega$ of the financial market. Let's assume that there are $n > 1$ assets in the market. The returns on these assets are represented by random variable $\tilde{r}: \Omega \to \mathbb{R}^n$ which has a normal distribution with the expected return $\boldsymbol{\pi}$ and the covariance matrix $\boldsymbol{\Sigma}$. That is

$$\tilde{r} \sim N(\boldsymbol{\pi}, \boldsymbol{\Sigma}) . \tag{1}$$

The BML uses "equilibrium" returns as a neutral starting point. Equilibrium returns are the set of returns that clear the market. The equilibrium returns are derived using a reverse optimization method in which the vector $\hat{\boldsymbol{\pi}} \in \mathbb{R}^n$ of implied excess equilibrium returns is extracted from known information using formula

$$\hat{\boldsymbol{\pi}} = \lambda \cdot \boldsymbol{\Sigma} \cdot \boldsymbol{w} \tag{2}$$

where $\lambda \in \mathbb{R}$ is the risk aversion coefficient and $\boldsymbol{w} \in \mathbb{R}^n$ is the vector of market capitalization assets weights. The risk-aversion coefficient characterizes the expected risk-return tradeoff. It is the rate at which an investor will forego expected return for less variance. In the reverse optimization process, the risk aversion coefficient acts as a scaling factor for the reverse optimization estimate of excess returns; the weighted reverse optimized excess returns equal the specified market risk premium.

More often than not, investment managers have specific views regarding the expected return of some of the assets in a portfolio, which differ from the implied equilibrium return.

In addition to the CAPM prior, the investor also has $k \geq 1$ views on the market returns. Any view is expressed as a statement that for fixed $i \leq k$ the linear combination of returns

$$\tilde{v}_i = \boldsymbol{p}_i^T \cdot \tilde{r} \tag{3}$$

has a normal distribution with the expected value $\varpi_i$ and the standard deviation $\varsigma_i$. The confidence to the view $\tilde{v}_i$ decreases with increase in standard deviation $\varsigma_i$. Then the investor's views can be expressed as system of linear equations

$$\boldsymbol{P} \cdot \tilde{r} = \tilde{v} \tag{4}$$

where

$$\tilde{v} \sim N(\varpi, \Xi) \tag{5}$$

and

$$P = [p_1^T, p_2^T, \ldots, p_k^T], \quad \tilde{v} = (\tilde{v}_1, \tilde{v}_2, \ldots, \tilde{v}_k)^T,$$

$$\varpi = (\varpi_1, \varpi_2, \ldots, \varpi_k)^T, \quad \Xi = \begin{bmatrix} \varsigma_1^2 & 0 & \cdots & 0 \\ 0 & \varsigma_2^2 & \cdots & 0 \\ \cdots & \cdots & \ddots & \cdots \\ 0 & 0 & \cdots & \varsigma_k^2 \end{bmatrix} \tag{6}$$

Taking into account a prior returns and additional investor's views we can obtain a posterior return having a normal distribution with the expected return $\pi_{BL}$ and the covariance matrix $\Sigma_{BL}$. That is

$$\tilde{r}_{BL} \sim N(\pi_{BL}, \Sigma_{BL}) \tag{7}$$

where

$$\Sigma_{BL} = ((\tau \cdot \Sigma)^{-1} + P^T \cdot \Xi^{-1} \cdot P)^{-1} \tag{8}$$

$$\pi_{BL} = \Sigma_{BL} \cdot ((\tau \cdot \Sigma)^{-1} \cdot \pi + P^T \cdot \Xi^{-1} \cdot \varpi) \tag{9}$$

for fixed scalar $\tau \in \mathbb{R}^+$. Walters [27] says that the meaning and impact of the parameter τ causes a great deal of confusion for many users of the BLM. Nevertheless, we can say that he confidence to the prior expected return $\pi$ versus investor's views decreases with increase in the parameter τ.

The two parameters of the BLM that control the relative importance placed on the equilibrium returns versus the investor's views, the scalar τ and the covariance matrix $\Xi$, are very difficult to specify. Litterman with the "Quantitative Resources Group, Goldman Sachs Asset Management" [16] point out that, "how to specify standard deviations $\varsigma_i$" is common question without a "universal answer". Regarding $\Xi$, Herold [12] says that the major difficulty of BML is that it forces the user to specify a probability density function for each view, which makes BLM only suitable for quantitative managers.

## 3. Intuitionistic fuzzy sets in the real line - basic concepts

Let us consider the space of all real numbers $\mathbb{R}$. The basic tool for imprecise classification of real numbers is the concept of fuzzy set $A \subset \mathbb{R}$ in which may be described as the set of ordered pairs

$$A = \{(x, \mu_A(x)): x \in \mathbb{R}\}. \tag{10}$$

where $\mu_A: \mathbb{R} \to [0,1]$ is its membership function. Intuitionistic fuzzy set [1] (for short IFS) $A \subset \mathbb{R}$ is defined as the set of ordered triples

$$A = \{(x, \mu_A(x), \nu_A(x)): x \in \mathbb{R}\}, \tag{11}$$

where nonmembership function $\nu_A: \mathbb{R} \to [0,1]$ fulfills the condition

$$\nu_A(x) \leq 1 - \mu_A(x) \tag{12}$$

for each $x \in \mathbb{R}$. The family of all IFS in the real line $\mathbb{R}$ we denote by symbol $\mathcal{I}(\mathbb{R})$.

We define hesitation function $\pi_A: \mathbb{R} \to [0,1]$ determined by the identity

$$\pi_A(x) = 1 - \mu_A(x) - \nu_A(x). \tag{13}$$

Value $\pi_A(x)$ indicates the degree of our hesitation in assessment of the relationship between the real number $x \in \mathbb{R}$ and IFS $A$. For this reason, the hesitation function $\pi_A$ may be interpreted as a image of knightian uncertainty [14].

For any $A, B \in \mathcal{I}(\mathbb{R})$ set theory operations are defined in the following way

$$A^C = \{(x, \nu_A(x), \mu_A(x)): x \in \mathbb{R}\}, \tag{14}$$

$$A \cup B = \{(x, \mu_A(x) \vee \mu_B(x), \nu_A(x) \wedge \nu_B(x)): x \in \mathbb{R}\}, \tag{15}$$

$$A \cap B = \{(x, \mu_A(x) \wedge \mu_B(x), \nu_A(x) \vee \nu_B(x)): x \in \mathbb{R}\}. \tag{16}$$

Let us consider fuzzy subset $B$ described by its membership function $\mu_B: \mathbb{R} \to [0,1]$. This fuzzy subset can be identified with IFS represented by the set of ordered triples

$$B^* = \{(x, \mu_B(x), 1 - \mu_B(x)); x \in \mathbb{R}\}, \tag{17}$$

The hesitation function of the above IFS identically fulfills the condition

$$\pi_B(x) = 0. \tag{18}$$

It implies that the fuzzy sets application to create real object model is implicit acceptance of strong assumption proclaiming that we are always able to decide on the fulfillment by each elementary state requirements postulated to its. As we know from everyday observations, however, usually it is not, and our settlements are burdened with a noticeable hesitation margin. This means that the extension of the fuzzy sets class to IFS class extends the capabilities of a reliable imprecision description.

IFS's are applied for description imprecise information's under knightian uncertainty. Many this subject researchers (e.g. [13]) distinguish two components of imprecision. They say that in the general case imprecision is composed of ambiguity and indistinctness. The information ambiguity is interpreted as a lack of clear recommendation one alternative between the various given alternatives. The information indistinctness we interpret, as the lack of explicit distinguishing amongst the given information and its negation. The hesitation function describes information insolubility which is interpreted as the lack of possibility to decide on the fulfillment by each elementary state requirements postulated to its. This insolubility causes knightian uncertainty.

Intensification of the information imprecision or information insolubility decreases this information usefulness. This gives rise to the problem of these phenomena evaluation. In this paper we use the following measure suggested in [21].

The ambiguity is evaluated by energy measure $d: \mathcal{I}(\mathbb{R}) \to [0; 1]$ given by the identity

$$d(A) = \lim_{y \to +\infty} \frac{\int_{-y}^{y} \mu_A(x) dx}{1 + \int_{-y}^{y} \mu_A(x) dx}. \tag{19}$$

The indistinctnees is measured by the most popular entropy $e: \mathcal{I}(\mathbb{R}) \to [0; 1]$ which is defined by Kosko [15] in the following way

$$e(A) = \frac{d(A \cap A^C)}{d(A \cup A^C)}.  \qquad (20)$$

The insolubility s evaluated by the ignorance measure $k: \mathcal{I}(\mathbb{R}) \to [0; 1]$ given by the identity

$$k(A) = d(A^*) - d(A_*)  \qquad (21)$$

where according to [2], for any IFS $A \in \mathcal{I}(\mathbb{R})$ we have

$$A^* = \{(x, \mu_A(x), 1 - \mu_A(x)): x \in \mathbb{R}\},  \qquad (22)$$

$$A_* = \{(x, 1 - \nu_A(x), \nu_A(x)): x \in \mathbb{R}\}.  \qquad (23)$$

An increase in imprecision or in insolubility significantly worsens the information quality. Thus using the vector-valued function $(d(\cdot), e(\cdot), k(\cdot))$ facilitates information quality management. Here it is desirable to minimize value of each coordinate.

## 4. Intuitionistic fuzzy posterior return

Let us reconsider investor's views which are additional information in BML. In the Section 2 each investor's view is represented by a random variable. It is obvious that the probability distribution of any investor's view is unobservable. Thus we can say that each investor's view is under knightian uncertainty. It implies that:

- Any investor's view cannot be represented by random variable.
- Any investor's view may be represented by IFS in the real line.

Therefore, in (3) random variable $\tilde{v}_i$ should be replaced by the IFS $V_i \in \mathcal{I}(\mathbb{R})$. Then we obtain the following condition

$$V_i = \boldsymbol{p}_i^T \cdot \check{\boldsymbol{r}}(\omega)  \qquad (24)$$

where $\check{\boldsymbol{r}}(\omega) = (\check{r}_1(\omega), \check{r}_2(\omega), \ldots, \check{r}_n(\omega))^T$ is a return determined for fixed elementary state $\omega \in \Omega$. It is obvious that any coordinate $\check{r}_i(\omega)$ is not a real number. Thus the vector $\check{\boldsymbol{r}}(\omega)$ is not a realization of random variable. IFS $V_i$ membership function describes the distribution of possible value of investor's views. IFS $V_i$ nonmembership function describes the distribution of impossible value of investor's views. Importance of each investor's view depends on this view usefulness. Thus importance of the investor's view is evaluated by means of the vector $(d(V_i), e(V_i), k(V_i))$. The view importance decreases with increase in any coordinate of this vector. The investor's view may not be an intuitionistic fuzzy number [9]

For example, the IFS $V_i$ may be given as expected return rates dependent on expected future value and intuitionistic fuzzy present value [21]. Moreover intuitionistic fuzzy present value can be determined as behavioural present value [20] which explicitly depends on observed market price and on impact of market conditions on the investor's beliefs. All it proves that IFS $V_i$ can be strictly determined as value which is verifiable.

For fixed elementary state $\in \Omega$ , immediately from (24) we obtain the system of linear equations

$$\boldsymbol{P} \cdot \check{\boldsymbol{r}}(\omega) = \boldsymbol{V} \tag{25}$$

where $\boldsymbol{V} = (V_1, V_2, \ldots, V_k)^T \in [\mathcal{J}(\mathbb{R})]^k$. In [22] is shown, that the system of equations (25) has the solution. The solution uniqueness is not discussed in [22]. Therefore, let us consider general solution as the indexed family of particular solutions

$$\boldsymbol{R}^{(\lambda)}(\omega) = \left(R_1^{(\lambda)}(\omega), R_2^{(\lambda)}(\omega), \ldots, R_n^{(\lambda)}(\omega)\right)^T \in [\mathcal{J}(\mathbb{R})]^n, \tag{26}$$

where $\lambda \in \Lambda$. Each IFS $R_i^{(\lambda)}(\omega)$ is represented by its conditional membership function $\rho_i^{(\lambda)}(\cdot | \omega): \mathbb{R} \to [0; 1]$ and its conditional nonmembership function $\varphi_i^{(\lambda)}(\cdot | \omega): \mathbb{R} \to [0; 1]$.

Let us consider now indexed family

$$\check{R}_i^{(\lambda)} = \left\{ R_i^{(\lambda)}(\omega): \omega \in \Omega \right\}. \tag{27}$$

which is a one alternative of posterior return on asset indexed by $i < n$ . This posterior return is the intuitionistic fuzzy probabilistic set [19] represented by its membership function $\rho_i^{(\lambda)}: \mathbb{R} \times \Omega \to [0; 1]$ determined by the identity

$$\rho_i^{(\lambda)}(x, \omega) = \rho_i^{(\lambda)}(x | \omega) \tag{28}$$

and by its nonmembership function $\varphi_i^{(\lambda)}: \mathbb{R} \times \Omega \to [0; 1]$ determined by the identity

$$\varphi_i^{(\lambda)}(x, \omega) = \varphi_i^{(\lambda)}(x | \omega). \tag{29}$$

Let posterior return on asset indexed by $i \leq n$ be denoted by the symbol $\check{R}_i$ . The posterior return $\check{R}_i$ is equal to union of all its alternatives $\check{R}_i^{(\lambda)}$. Thus, his posterior return is represented by its membership function $\rho_i: \mathbb{R} \times \Omega \to [0; 1]$ determined by the identity

$$\rho_i(x, \omega) = \sup \left\{ \rho_i^{(\lambda)}(x, \omega): \lambda \in \Lambda \right\} \tag{30}$$

and by its nonmembership function $\varphi_i: \mathbb{R} \times \Omega \to [0; 1]$ determined by the identity

$$\varphi_i(x, \omega) = \inf \left\{ \varphi_i^{(\lambda)}(x, \omega): \lambda \in \Lambda \right\}. \tag{31}$$

Finally we obtain the posterior return given as vector

$$\check{\boldsymbol{R}} = \left( \check{R}_1, \check{R}_2, \ldots, \check{R}_n \right)^T . \tag{30}$$

Immediately from (25) we obtain that the probability measure $\mathcal{P}: 2^\Omega \supset \sigma \to [0,1]$ is uniquely defined by the prior distribution that is CAPM equilibrium distribution.

In this way, we gathered all information necessary for described in [21] analysis of intuitionistic fuzzy return rate.

## 5. Final conclusions

In this paper BML is modified in this way that randomized investor's views are replaced by intuitionistic fuzzy views. This replacement is justified by means of the observation that investor's view are under knightian uncertainty. In this way we obtain the model independent on two parameters which control the relative importance placed on the equilibrium returns versus the investor's views, the scalar τ and the covariance matrix Ξ. Let us remind ourselves that the meaning and impact of the first parameter causes a great deal of confusion for many users of the BLM [27]. Moreover, in subject literature we cannot to find well justified method of covariance matrix Ξ estimation. Recapitulating, this parameters elimination allows us to replace the BML by modified BML which is free from subjective evaluations significance of investor views. This is the basic advantage of the proposed modifications BML.

Here it is the only proven that posterior return exists. Thus, the results so obtained may be only applied in finance theory as the normative model. On the other side, these results can be directly used in the decision making models described in [10]. It causes that results presented above can constitute a theoretical foundation for constructing investment decision support system.

Applications of the normative model presented above involve several difficulties. The main difficulty is the high formal and computational complexity of the tasks of determining the membership and nonmembership functions of posterior return. Computational complexity of the normative model is the price which we pay for the lack of detailed assumptions about investor's views. On the other hand, low logical complexity is an important good point of the formal model presented in this paper.

**Acknowledgements:** The project was supported by funds of National Science Center - Poland granted on the basis the decision number DEC-2012/05/B/HS4/03543.

## References

[1] Atanassov, K., S. Stoeva S.: Intuitionistic fuzzy sets, Proc. of Polish Symp. on Interval & Fuzzy Mathematics, Poznan, Aug. (1983), 23-26.

[2] Atanassov K.: New variant of modal operators in intuitionistic fuzzy modal logic, BUSEFAL 54 (1993), 79-83.

[3] Bevan A., Winkelmann K.: Using the Black-Litterman Global Asset Allocation Model: Three Years of Practical Experience. Goldman, Sachs & Co Fixed Income Research, (1998).

[4] Black, F.: Equilibrium Exchange Rate Hedging, NBER Working Paper Series: Working Paper No. 2947, (1998).

[5] Black F.: Universal Hedging: Optimizing Currency Risk and Reward in International Equity Portfolios, Financial Analysts Journal, 45 (1998), 16-22. DOI: 10.2469/faj.v45.n4.16

[6] Black F., Litterman R.: Asset allocation: Combining investor views with market equilibrium. The Journal of Fixed Income 01/1991; 1(2) (1990), 7-18. DOI: 10.3905/jfi.1991.408013

[7] Black F., Litterman R.: Global Asset Allocation with Equities, Bonds, and Currencies. Goldman, Sachs & Co Fixed Income Research, (1991).

[8] Black F., Litterman R.: Global Portfolio Optimization. Financial Analysts Journal, September/October, (1992), 28-43.

[9] Burillo P., H. Bustince H., Mohedano V.: Some definition of intuitionistic fuzzy number, Fuzzy based expert systems, fuzzy Bulgarian enthusiasts, Sofia, Sept. (1994), pp 28-30.

[10] Deng-Feng Li.: Decision and Game theory in Management with Intuitionistic Fuzzy Sets. Springer, Berlin - Heidelberg, (2014).

[11] He G., Litterman R.: The Intuition Behind Black-Litterman Model Portfolios, SSRN Electronic Journal (2002); DOI: 10.2139/ssrn.334304

[12] Herold U.: Portfolio Construction with Qualitative Forecasts, Journal of Portfolio Management 30 (2003), 61-72. DOI: 10.3905/jpm.2003.319920

[13] Klir G.J.: Developments in uncertainty-based information, Advances in Computers 36, (1993), 255-332.

[14] Knight F. H.: Risk, Uncertainty, and Profit, Hart, Schaffner & Marx; Houghton Mifflin Company, Boston, MA, (1921)

[15] Kosko B.: Fuzzy entropy and conditioning, Information Sciences 40, (1986), 165-174. DOI: 10.1016/0020-0255(86)90006-X

[16] Litterman R., the Quantitative Resources Group, Goldman Sachs Asset Management: Modern Investment Management: An Equilibrium Approach. New Jersey: John Wiley & Sons. (2003).

[17] Litterman R., Winkelmann K.: Estimating Covariance Matrices, Goldman, Sachs & Co Risk Management Series, (1998).

[18] Markowitz H.M.: Portfolio Selection, The Journal of Finance 7, (1952), 77-91.

[19] Qiansheng Zhang, Baoguo Jia, Shengyi Jiang: Interval-valued intuitionistic fuzzy probabilistic set and some of its important properties, Proceedings of the 1st International Conference on Information Science and Engineering ICISE 2009, Guangzhou, (2009), 4038 - 4041, DOI: 10.1109/ICISE.2009.692

[20] Piasecki K.: Intuitionistic assessment of behavioural present value, Folia Oeconomica Stetinensia 13(21) (2013), 49-62, DOI: 10.2478/foli-2013-0021

[21] Piasecki K. On return rate estimated by intuitionistic fuzzy probabilistic set,. In: D. Martincik, J. Ircingova, P. Janecek (Eds.). Mathematical Methods in Economics MME 2015. pp 641-646. Faculty of Economics, University of West Bohemian Plzen (2015)

[22] Pradham R., Pal M.: Solvability of system of intuitionistic fuzzy linear equations, International Journal of Fuzzy Logic Systems 4 (2014) 13-24.

[23] Sharpe W.F.: Capital Asset Prices: A Theory of Market Equilibrium, Journal of Finance 19, (1964), 425-442. DOI: 10.1111/j.1540-6261.1964.tb02865.x

[24] Sharpe W.F.,: Imputing Expected Security Returns from Portfolio Composition. Journal of Financial and Quantitative Analysis 9, (1974), 463-472. DOI: 10.2307/2329873.

[25] Theil H.: Principles of Econometrics. New York: Wiley and Sons, (1971).

[26] Theil H.: Introduction to Econometrics. New Jersey: Prentice-Hall, Inc. (1978)

[27] Walters J. The Black-Litterman model in detail, SSRN Electronic Journal (2011); DOI: 10.2139/ssrn.1314585